\documentclass{article}
\input{macros_COVID.sty}
\usepackage[colorinlistoftodos,bordercolor=orange,backgroundcolor=orange!20,linecolor=orange,textsize=scriptsize]{todonotes}

\title{Probabilistic and mean-field model of COVID-19 epidemics with user mobility and contact tracing}

\author{M. Akian\footnote{INRIA and CMAP, École polytechnique, IP Paris, CNRS.  Email: \texttt{marianne.akian@inria.fr}.}\, , L. Ganassali\footnote{INRIA, DI/ENS, PSL Research University, Paris, France. Email: \texttt{luca.ganassali@inria.fr}.}\, , S. Gaubert\footnote{INRIA and CMAP, École polytechnique, IP Paris, CNRS.  Email: \texttt{stephane.gaubert@inria.fr}.}\, , L. Massoulié\footnote{MSR-Inria Joint Centre, INRIA, DI/ENS, PSL Research University, Paris, France. Email: \texttt{laurent.massoulie@inria.fr}.}}
\date{\today}

\begin{document}

\maketitle
\begin{abstract}
  We propose a detailed discrete-time model of COVID-19 epidemics coming in two flavours, mean-field and probabilistic. The main contribution lies in  several extensions of the basic model that capture i) user mobility -- distinguishing {\em routing}, i.e.\ change of residence, from {\em commuting}, i.e.\ daily mobility -- and ii) contact tracing procedures. We confront this model to public data on daily hospitalizations, and discuss its application as well as underlying estimation procedures.
\end{abstract}

\tableofcontents

\section*{Introduction}

A profusion of mathematical models of the COVID-19 epidemic propagation has been developed in a very short time lapse. Specific models were designed to address particular objectives, such as forecasting the epidemics, understanding the impact of various interventions such as restraining mobility and contacts between users, or deploying contact tracing and case isolation. 

The goal of this study is to propose a single model of COVID-19 epidemics based on a well understood family of probabilistic models -- that of multi-type branching processes -- and to detail simple extensions of this model that capture the impact of contact tracing procedures as well as user mobility. 


Our model can serve several purposes: It can be used to assess from available time series so-called contact rates, and in turn the impact of user mobility on epidemic propagation. It can also be used in surveillance platforms to infer the status of specific sub-populations (in particular fraction of infectious or immunized persons), a prerequisite for the implementation of targeted testing campaigns.

\paragraph{Paper organization}
We introduce in Section \ref{section_model} a discrete time epidemiological model of COVID-19 for a homogeneous population. 
We detail a probabilistic version of the model, that is a multi-type branching process. It represents in a precise manner the durations of the principal phases of a patient's condition and the transition rates between such phases. A deterministic version of this model is then described. We then introduce a simple extension of the model to represent the impact of contact tracing and case isolation.

In Section \ref{section_inference}, we describe two methods that both use model structure for inference of parameters and prediction. For the deterministic, or mean-field version of the model, we propose an approach to perform inference that relies on Perron's theory. For the probabilistic version of the model, by exploiting the fact that we are dealing with a hidden Markov model, we propose a Kalman filtering approach to perform inference.

Next, we present a numerical illustration in Section \ref{section_numerical}, where the model is fitted to daily counts of hospitalizations in the Paris area. There we estimate piecewise constant contact rates corresponding to distinct time periods, namely pre-confinement, intermediate, and confinement periods. We discuss the application of this fitting procedure to forecasting the future number of hospitalizations. 


We then describe in Section \ref{section_mobilite} an extension for the contact rates to take daily mobility -- or other covariates -- into account, making these parameters depend on time.

We finally discuss in Section \ref{section_routing} the extension of the model to multiple sub-populations, corresponding e.g. to geographical regions and age ranges. Key parameters in this extension are contact matrices, describing the frequency of contacts between sub-populations. The routing dimension of this extension is detailed.  we revisit the question of estimating model parameters, specifically contact matrices in this multi sub-populations model. We assume potentially time-varying contact matrices and consider estimation of such contact matrices based on user mobility traces. 





\paragraph{Related work}
SIR or SEIR-type models of COVID-19 featuring asymptomatic or undocumented infectious individuals have been considered in \cite{lipeichen2020}, \cite{singh} and \cite{CS-Chang2020}. Studies \cite{singh,CS-Chang2020} consider same contact rates for symptomatic and asymptomatic cases. Some works \cite{CrepeyMassonaud2020} also consider a split of the infectious phase to explicit the occurrence of aggravated symptoms. 

General distributions of times spent in a given state have been studied.  Prodromic phase, during which individuals are infectious, yet have not displayed symptoms, is considered in \cite{colizza_R9}. The model of \cite{colizza_R9} comprises a detailed breakdown in various phases, yet relies on geometric/exponential durations for time spent in each phase, while the model of \cite{Sofonea2020} describes general discrete time distributions for the time spent in states.

A broad range of studies have addressed the issue of estimating the time spent in a given state. In \cite{McAloon20}, the incubation period is modeled with a log-normal distribution with mean 5.8 days. In \cite{Lauer20}, the median incubation period was estimated to be 5.1 days, and $97.5\%$ of those who develop symptoms will do so within 11.5 days of infection.

\cite{Byrne20} reviews studies of the infectious period's duration, reporting several estimates with substantial variations, giving for asymptomatic cases an estimate of 6.5-9.5 days. Pre-symptomatic infectious period is estimated to 0-4 days across studies. \cite{Verity20} also investigates the time between infection and remission/death.

The model of \Cref{fig_SEIAR_model_1} is a simplified version of the model considered in \cite{colizza_R9}, which features additional states, including a breakdown between hospitalized cases entering Intensive Care Units or not, and a distinction between several severity levels of the infection. It is similar to the model considered in \cite{CrepeyMassonaud2020}, with a key
difference in our model that is that we take the discrete time $t$ into
account.
\section{Single Population Model}
\label{section_model}
\begin{figure}[H]
\centering
\includegraphics[width=0.95\textwidth]{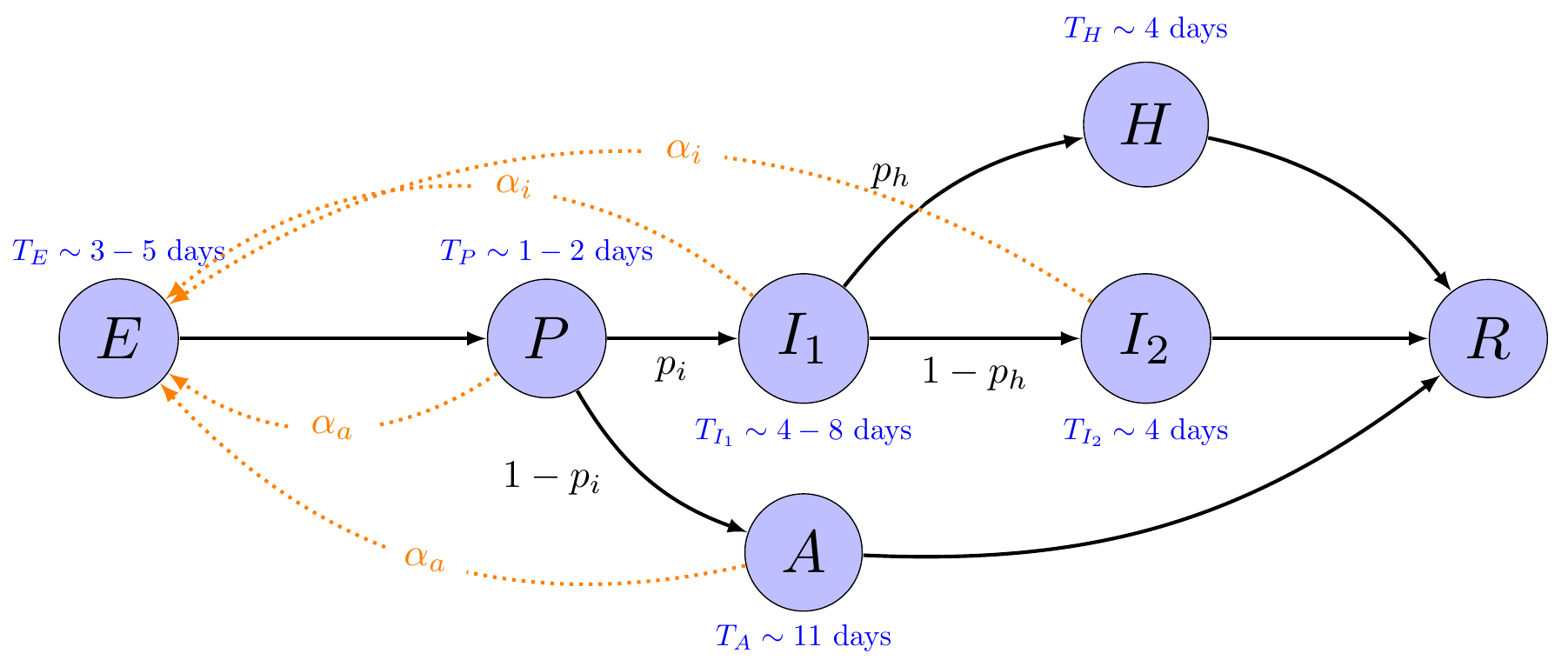}
\caption{State diagram for individual's condition in model 1. Dotted arrows show the contamination of new cases: the mean number of new $E$ cases per day per $A$/$P$ (resp. per $I_1$/$I_2$) individual is $ \alpha_a$ (resp. $ \alpha_i$).}
\label{fig_SEIAR_model_1} 
\end{figure}
We consider the discrete time SEIR-type model depicted in \Cref{fig_SEIAR_model_1}. We classify the states of an individual as $E$ for {\em exposed}; $P$ for {\em prodromic}; $A$ for {\em infectious asymptomatic}; $I_1$ for {\em phase 1 infectious symptomatic}; $H$ for {\em hospitalized}; $I_2$ for (non-hospitalized) {\em phase 2 infectious symptomatic};
and $R$ for {\em removed} from the contamination chain. After having moved to prodromic phase $P$, an individual becomes either asymptomatic (state $A$) or symptomatic, entering state $I_1$. After some time in the $I_1$ state, two situations arise: either the individual is hospitalized, entering state $H$, either there is no aggravation, leading to state $I_2$. Then, the last state is always $R$, where the individual is either dead  or recovered.\\

We shall in fact adopt a more detailed representation for the state of an individual: its detailed state will be represented by a couple $(a,t)$ where $a\in \{E,A,I_1,H,I_2,R\}$ is the stage of the disease, and $t$ the number of days spent in this stage. For instance, an individual in state $(E,3)$ just completes their third day in the exposed phase.\\


\begin{remark}
\begin{itemize}
\item[$(i)$] In epidemiological models one often tracks individuals in so-called {\em susceptible } state $S$: these are those individuals who may get exposed to infection. In the case of a negligible fraction of infected individuals, in particular in the early stages of the epidemics, on which we focus, it is not necessary to track them, hence our choice not to include state $S$ in the diagram of \Cref{fig_SEIAR_model_1}.

\item[$(ii)$] We distinguish between asymptomatic and symptomatic individuals, as new symptomatic cases can be observed when the corresponding individuals consult their physician or enter hospital after experiencing serious symptoms.  
\item[$(iii)$] We distinguish between hospitalized and non-hospitalized cases.
  The number of new hospital admissions can be measured, for instance, from medical emergency services records.
\item[$(iv)$] We distinguish according to the number of days an individual has spent in its current state for the following reason. Traditional SEIR-type models implicitly assume that the time that an individual spends in a particular state admits an exponential distribution (for continuous time ODE-based models) or a geometric distribution (for discrete time models). However the durations spent in the states of interest for COVID-19 appear to be very far from exponentially distributed. Thus, for an accurate short-term prediction of the epidemics evolution, one needs to take this number of days into account.
\end{itemize}
\end{remark}

\paragraph{State variables}
We  represent the dynamics by means of the following state variables.
On each day $t$, we represent the population
by a vector $x(t)$ with entries:
\begin{flalign*}
x_{E,d}(t):&\hbox{ number of individuals in phase $E$, having spent $d$  days in this phase};\\
x_{P,d}(t):&\hbox{ number of individuals in phase $P$, having spent $d$  days in this phase};\\
x_{I_1,d}(t):& \hbox{ phase 1 infectious individuals, having spent $d$  days in this phase};\\
x_{I_2,d}(t):& \hbox{ phase 2 infectious individuals having spent $d$  days in this phase};\\
x_{A,d}(t):& \hbox{ asymptomatic individuals having spent $d$  days in this phase};\\
x_{H}(t):& \hbox{ number of individuals newly hospitalized on day } t.
\end{flalign*}
By considering the state variable $x_H(t)$, instead of $x_{H,d}(t)$,
we ignore the number of days an individual spends at hospital.
This stems from the assumption that such cases are efficiently isolated and do not contribute to propagating the epidemics, an assumption that could be revised.

\paragraph{Distribution of phase durations}
For each $\tau \in \{E,P,I_1,I_2,A\}$, define $r_\tau(d)$ the probability that corresponding phase will end on the following day, given that it has lasted $d$ days. Denoting by $p_\tau(d)$ the probability that phase $\tau$ lasts $d$ days, $r_\tau(d)$ is the associated failure rate, defined by
\begin{equation}
r_\tau(d)=\left\{
\begin{array}{ll}
\frac{p_\tau(d)}{\sum_{\delta\ge d} p_\tau(\delta)}&\hbox{ if }\sum_{\delta=1}^d p_\tau(\delta)>0,\\
0&\hbox{ otherwise.}
\end{array}
\right.
\end{equation} Available statistics suggest that typical durations lie in the ranges given on \Cref{fig_SEIAR_model_1}. A reasonable baseline assumption could consist in taking
\begin{itemize}
\item $p_E(d)=1/3$ for $d\in\{3,4,5\}$ and $p_E(d)=0$ otherwise,
\item $p_P(1)=p_P(2)=1/2$ and $p_{P}(d)=0$ otherwise,
\item $p_{I_1}(d)=1/3$ for $d\in\{5,6,7\}$ and $p_{I_1}(d)=0$ otherwise,
\item $p_{I_2}(d)=1$ for $d=4$ and $p_{I_2}(d)=0$ otherwise,
\item $p_A(d)=1$ for $d=11$ and $p_{A}(d)=0$ otherwise.
\end{itemize} Note however that far longer durations have been observed, e.g. incubation lasting $10$ days or more; $I_1$ phase between first symptoms and aggravation lasting as long as $20$ days. This baseline is therefore by no means an accepted characterization of the dynamics of an individual's condition.

\paragraph{Transition probabilities}
\begin{itemize}
\item $p_i$: probability that previously exposed individual becomes symptomatic at end of incubation. A baseline assumption consists in taking $p_i=0.7$. Again, no consensus has emerged yet on the fraction of asymptomatic infected users and competing assumptions have been proposed (for baselines, data from APHP \textit{https://www.aphp.fr} can be used, as done e.g. in \cite{colizza_R9}).
\item $p_h$: probability that previously phase 1 infectious individual develops aggravated form at end of phase 1. A baseline assumption consists in taking $p_{h}=0.05$, again with the same cautionary note. 
\end{itemize}

\paragraph{Contact rates}
\begin{itemize}
\item $\alpha_i$: average number of new exposed individuals generated by single infectious individual ($I_1, I_2$) in a single day. Here we do not distinguish between phases 1 and 2, an assumption which could be revisited.

\item $\alpha_a$: average number of new exposed individuals generated by single asymptomatic or prodromic individual in a single day. 
\end{itemize} In the sequel we assume that $\alpha_a=\alpha_i$. This assumption is often made, for instance in \cite{singh,CS-Chang2020}. In contrast, \cite{lipeichen2020} makes an explicit distinction in contact rates $\alpha$ for documented cases and $\mu \alpha$ for undocumented ones. Especially, an estimate for $\mu$ in China between January 10th and January 23th is given: $\mu \sim 0.55$ ($95\%$ confidence interval: $(0.46, 0.62)$).

\begin{remark}
This model can be complemented by equations giving the number of deaths and immunizations per day. This, together with a tracking of the volume of compartment $S$, could be useful if one wishes to estimate the proportion of immunized individuals in the population. Denoting by $p_d$ the probability of death conditional on having been hospitalized, and assuming that only hospitalized individuals die, this leads to two other state variables:
\begin{flalign*}
x_{\mathrm{death}}(t+1)&=p_d x_H(t),\\
x_{\mathrm{immun}}(t+1)&=(1-p_d) x_H(t)+\sum_{\delta>0}x_{A,\delta}(t)r_A(\delta)+x_{I_2,\delta}(t)r_{I_2}(\delta).
\end{flalign*}The number of deaths as written neglects time between hospitalization and death.
\end{remark}

\paragraph{Observables}
We assume that coordinates $x_H(t)$ are observed by the end of day $t$. This would correspond to the number of new hospitalized patients with COVID-19 pathology. In contrast, the states variables $x_E(t), x_A(t), x_{P}(t)$ are never observed. They will later be referred to as \textit{hidden trajectories}. It would be straightforward to adjust the model to incorporate other observables, such as the number of patients consulting their physician on each day. 

\subsection{Deterministic version: mean-field model}
\label{subsection_mean_field_model}
This is a deterministic, linear model, based on the assumption that the fraction of susceptible individuals in the total population is close to 1. In other words it is meant to represent stages of the epidemics where collective immunization is yet negligible. Simple adjustments can be made to transform this model to account for a non-negligible, time-evolving, fraction of immunized population. 

Assume there is a maximal number of days, $h$, that each of the phases $E$, $P$, $I_1$, $I_2$, $A$ can last. Let $x_E=(x_{E,1},\ldots,x_{E,h})^\top$,  $x_{P}=(x_{P,1},\ldots,x_{P,h})^\top$, and similarly for $h$-dimensional vectors $x_A$, $x_{I_1}$, $x_{I_2}$. The state space on day $t$, $x(t)$, is given by

\begin{equation}
x(t) := \begin{pmatrix}
x_E(t)\\
x_P(t)\\
x_{I_1}(t) \\
x_{A}(t) \\
x_{I_2}(t) \\
x_{H}(t) \\
\end{pmatrix}
\end{equation}The dynamics are then given by
\begin{equation}
\label{eq_matrix_M}
x(t+1)=M x(t),
\end{equation} where matrix $M$ is specified by the detailed equations, for $d\in[h-1]$:

\begin{flalign*}
\forall \tau \in \left\lbrace E,P,I_1,I_2,A \right\rbrace, \;  x_{\tau,d+1}(t+1)&= x_{\tau,d}(t)(1-r_{\tau}(d)),\\
x_{E,1}(t+1) &=\sum_{\delta>0}\left[\alpha_i(x_{I_1,\delta}+x_{I_2,\delta})(t)+\alpha_a ( x_{A,\delta}+ x_{P,\delta})(t)\right] ,\\
x_{P,1}(t+1)&=\sum_{\delta>0} x_{E,\delta}(t) r_E(\delta),\\
x_{I_1,1}(t+1)&=p_{i}\sum_{\delta>0}x_{P,\delta}(t)r_P(\delta),\\
x_{A,1}(t+1)&=(1-p_{i})\sum_{\delta>0}x_{P,\delta}(t)r_P(\delta),\\
x_{I_2,1}(t+1)&=(1-p_{h})\sum_{\delta>0}x_{I_1,\delta}(t) r_{I_1}(\delta),\\
x_{H}(t+1)&=p_{h}\sum_{\delta>0}x_{I_1,\delta}(t)r_{I_1}(\delta).
\end{flalign*} 

\subsection{Probabilistic version: multi-type branching process}
\label{subsection_branching_model}
\begin{figure}[H]
	\centering
	\includegraphics[width=\textwidth]{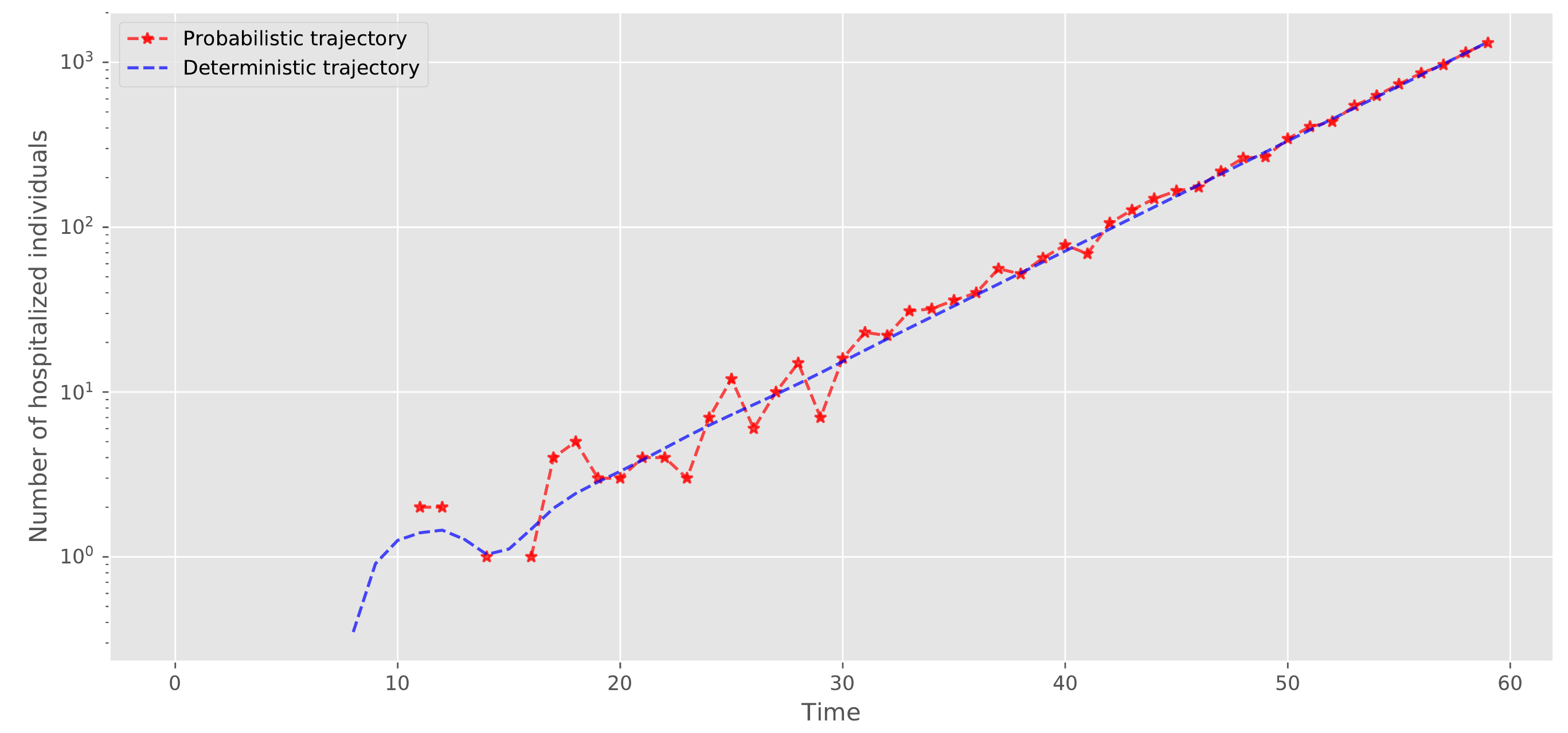}
	\caption{Trajectories of newly hospitalized individuals of length $60$ days from the probabilistic and deterministic versions, with $p_a = 0.3$, $p_h = 0.05$, $\alpha_i = 0.4$, $\alpha_a = 0.3$, initialized with $200$ individuals entering in $E$.}
	\label{fig_traj_det_VS_prob}
\end{figure}

We now transform the previous mean field model into a probabilistic counterpart: a multi-type (or multi-dimensional) branching process. Consideration of probabilistic behaviour is particularly relevant when we deal with not-so-large infected populations, or in sub-critical cases where the largest eigenvalue $\lambda_1(M)$ of matrix $M$ defined in \eqref{eq_matrix_M} verifies $\lambda_1(M)\le 1$. Indeed, when $\lambda_1(M)>1$, by results of Kesten and Stigum \cite{MR0198552}, there is a random variable $Z$ such that for large $t$, the state $X(t)$ renormalized by $\lambda_1(M)^{-t}$ converges to $Z u$, where $u$ is the eigenvector of $M$ associated with eigenvalue $\lambda_1(M)$. This provides a theoretical justification of the mean field model in the case of large infected populations. \Cref{fig_traj_det_VS_prob} gives a numerical illustration of this result.\\

The number of newly exposed individuals now follows a Poisson distribution with mean as specified in the mean field model. Determination of durations in each state, and choice of next state at branches in the state diagram of \Cref{fig_SEIAR_model_1}, are made with independent random coins with the prescribed probabilities.

For clarity, to distinguish between the two versions, random variables will be denoted in capital letters through the rest of this article. Binomial (resp. Poisson) distributions are denoted by $\cB$ (resp. $\cP$).

Thus, conditioning to state vector $X(t)$ at time $t$, the state $X(t+1)$ at time $t+1$ verifies:
\begin{flalign*}
\forall \tau \in \left\lbrace E,P,I_1,I_2,A \right\rbrace, \;  X_{\tau,d+1}(t+1)& \sim \cB  \left(X_{\tau,d}(t), 1-r_{\tau}(d)\right),\\
X_{E,1}(t+1) & \sim \cP \left( \sum_{\delta>0}\left[\alpha_i(X_{I_1,\delta}+X_{I_2,\delta})(t)+\alpha_a (X_{A,\delta}+ X_{P,\delta})(t) \right] \right) ,\\
X_{P,1}(t+1)&=\sum_{\delta>0} \left[ X_{E,\delta}(t) - X_{E,\delta+1}(t+1) \right],\\
X_{I_1,1}(t+1)& \sim \cB\left(\sum_{\delta>0}\left[X_{P,\delta}(t)-X_{P,\delta+1}(t+1)\right],p_i\right),\\
X_{A,1}(t+1)&=\sum_{\delta>0}\left[X_{P,\delta}(t)-X_{P,\delta+1}(t+1)\right] - X_{I_1,1}(t+1),\\
X_{I_2,1}(t+1)& \sim \cB\left( \sum_{\delta>0}[X_{I_1,\delta}(t) - X_{I_1,\delta+1}(t+1)],1-p_h \right),\\
X_{H}(t+1)&=\sum_{\delta>0}\left[X_{I_1,\delta}(t) - X_{I_1,\delta+1}(t+1)\right] - X_{I_2,1}(t+1).
\end{flalign*} 

\subsection{Extension with contact tracing} 
\label{subsection_contact_tracing}
One motivation of the present study is the control of the epidemics via case isolation, that is quarantining of identified cases so that they do not contribute to further virus propagation. The main approach to deploy case isolation relies on identification of infectious cases via tests. Thus, when testing capacity is limited, efficient case isolation crucially depends on targeted test campaigns. Contact tracing is of paramount importance for targeting tests: it allows to determine who has been in contact with infectious cases, and hence focus tests on such contacts.\\


We now describe an extension of our previously introduced model to include contact tracing and case isolation. In a nutshell, the model assumes that those individuals subsequently infected by a specific person (its children in the branching tree associated to the branching process) get tested with some probability once that specific person is positively tested (and subsequently isolated). 
This formalization of contact tracing has already been considered in \cite{Lambert2020} for a simple branching process, analyzing the efficiency of contact tracing plus case isolation to control epidemics. We give here a more general version,  providing a generic method to expand the state space of a multi-type branching process in order to capture the introduction of contact tracing plus case isolation.\\


Consider then a multi-type branching process where user types are indexed by $j\in[J]$. For the sake of generality, we shall assume that types $j$ encode the succession of future phases $\phi$ that an individual will visit over the coming days. Thus an individual with type $j=\{E,E,P,I_1,H\}$ is in state $E$ for the present day and the next, and will then spend a single day in the states $P$, then $I_1$, and then $H$ subsequently. Also, an individual with type $j=\{\phi_1,\ldots,\phi_\ell\}$ on a given day will become an individual with type $j'=\{\phi_2,\ldots,\phi_{\ell}\}$ on the next day. 

Note that the previous branching process gives rise to an equivalent branching process with this set of types, as can be seen by sampling, upon birth of an individual, the sequence of states $\phi$ it will visit after its birth.

We now assume that there is a set $\Phi_0$ of distinguished states $\phi\in \Phi_0$ such that an individual entering  a state $\phi$ in $\Phi_0$ automatically incurs a {\em positive test}. Concretely, the state $\phi=H$, namely hospitalization of the individual, would automatically trigger a positive test, and we may think of $\Phi_0$ as the singleton $\{H\}$.

We then assume that, for some fixed tracing probability $p_t>0$, the contact between an infectious individual and another individual that became infected is being recorded. The original branching process dynamics are then modified as follows: once an individual is tested positive, it is isolated (removed) and its traced contacts are subsequently tested positive on the following day, and subsequently isolated, and so on. 

The key observation is that this modification preserves the branching structure, provided we perform the following extension of the original type space $[J]$. The new type space becomes 
$$
\mathcal{J}=\{(j,d):j\in[J],d\in [D_{\max}]\cup\{+\infty\}\},
$$
where the second coordinate $d$ represents the number of days until the corresponding individual will be positively tested.

Upon birth of an individual whose parent has type $(j,d)$, the newly born individual will have type $(j',d')$. We now specify the distribution of the pair $(j',d')$. First sample a sequence $j''=\{\phi''_1,\ldots,\phi''_{\ell''}\}\in [J]$ conditionally on $j$ with the same conditional distribution as in the original branching process. Let $d_1=\inf\{k\ge 1: \phi'_k\in\Phi_0\}$ be the number of days before the newly born individual enters some state in $\Phi_0$. 

We then let 
\begin{equation}
d'=\left\{
\begin{array}{ll} d_1&\hbox{with probability }1-p_t,\\
\min(d_1,d+1)&\hbox{ with probability }p_t.
\end{array} \right.
\end{equation}
this reflects the fact that if traced, the new individual will become tested either when it enters some state in $\Phi_0$, or on the day after its parent gets tested, whichever happens first. If untraced, it only becomes tested on the day it enters some state in $\Phi_0$. Finally, we let $j'=\{\phi''_1,\ldots,\phi''_{d'}\}$, i.e.\ we truncate the sequence $j''$ to length $d'$, thereby encoding the impact of case isolation, by which a positively tested individual can no longer infect anyone else.\\

Note that further extensions can be considered that preserve the branching process structure. Here is an example of such an extension. Assume that on each day, each infected individual may be submitted to a randomly administered test, and trigger a positive test reply, this happening independently with probability $\eps$ on any given day. To reflect such random tests, the above-described dynamics would then need to be amended in the following manner. For an individual newly born from a $(j,d)$-parent, sample a random variable $X$ distributed as 
\begin{equation}
\label{eq_d_distri}
X = G \cdot \mathbf{1}_{G \leq D_{\max}} + \infty \cdot \mathbf{1}_{G>D_{\max}},
\end{equation} where $G$ follows a geometric distribution of parameter $\eps$.\\

With $j''$ and $d_1$ as above, let then 

\begin{equation}
d'=\left\{
\begin{array}{ll} \min(d_1,X)&\hbox{with probability }1-p_t,\\
\min(d_1,X,d+1)&\hbox{ with probability }p_t.
\end{array} \right.
\end{equation}
The construction then goes through unchanged, letting $j'=\{\phi'_1,\ldots,\phi''_{d'}\}$ and the newly born individual's type being $(j',d')$.\\

The above extensions to the original branching process can be used  to assess how the tracing probability $p_t$, or the fraction $\epsilon$ of random tests, affects the potency of the epidemics. In particular, one may consider  the mean progeny matrix $M$, indexed by extended types $j\in[J']$ and associated with the branching process that captures contact tracing. 

A crude criterion for success of contact tracing consists in requiring that the spectral radius $\rho(M)$ be strictly less than 1. Under such conditions, given a vector $x(0)$ capturing the initial number of individuals of each type, then the expectation of $X_{\mathrm{infected}}$, the total number individuals ultimately infected, reads
\begin{equation}
\dE (X_{\mathrm{infected}})=\sum_{t\ge 0} \langle x(0), M^t e\rangle=\langle x(0), (I-M)^{-1}e\rangle.
\end{equation}
In the above, $e$ denotes the all-ones vector and the second equality is valid whenever $\rho(M)<1$. 

This expression for $\dE (X_{\mathrm{infected}})$ can be further processed to obtain simple upper bounds. Particularly simple upper bounds have been obtained in \cite{draief2008} in the case where $M$ is symmetric. However symmetry does not hold for the matrices $M$ considered here, so that we cannot rely on the bounds of \cite{draief2008}.


\section{Inference methods}
\label{section_inference}
In this section, we propose inference methods for our model. We consider a general deterministic (respectively, probabilistic) setting where the state vector at time $t$ is denoted by $x(t)$ (resp. $X(t)$). 

Individual types are now denoted by $j \in [J]$, and $M_{i,j}$ is the mean number of type $i-$children from a type $j-$parent. $M^\top$ is known as the \textit{mean-progeny matrix} in branching process theory. It depends on a parameter of interest $\theta \in \Theta$, that encapsulates for instance $p_i, p_h, \alpha_a, \alpha_i$ and time distributions in Section \ref{section_model}, as well as $p_t$ and $c_\mathrm{max}$ if we consider contact tracing as in Section \ref{subsection_contact_tracing}. 

We assume that the observed trajectories are the first $h$ coordinates of $x(t)$ (resp. $X(t)$), denoted by $x_H(t)$ (resp. $X_H(t)$), all other coordinates being hidden trajectories.

\subsection{Perron's theory for fitting and inference in the mean-field setting}
\label{subsection_perron}
For simplicity, assume that the remaining unknown parameters of the model are the contact rates $\alpha_a$ and $\alpha_i$, all other parameters having correctly been estimated, e.g. from patient statistics compiled at hospitals. 

\paragraph*{Constant contact rates}
In the case where contact rates are constant at $\theta_0 = (\alpha_{i,0},\alpha_{a,0})$, the state at time $t$ is given by $x(t)=M(\theta_0)^tx(0)$. Its behavior is determined at first order by the largest eigenvalue of matrix $M=M(\theta_0)$,  $\lambda_1(M)$, $x(t)$ being close to $c \lambda_1(M)^t u$, where $u$ is the eigenvector of $M$ associated with eigenvalue $\lambda_1(M)$ and $c$ some constant. The exponential rate of growth $\lambda_1(M)$ of observable $x_H(t)$ can be estimated, e.g. by fitting a regression line to the semilog plot up to time $t$ of the corresponding time series, thus producing the estimate $\hat{\lambda}(t)$. 


For  a single population, in the mean field model, estimation of the exponent $\lambda_1(M)$ gives one relation constraining the unknown parameters, namely:
\begin{equation}
\hat{\lambda}(t)=\lambda_1(M(\theta_0))=:F(\theta_0).
\end{equation}

\begin{remark}\label{rem_ar} Let $p$ be a polynomial such that $p(M)=0$ ($p$ could be the characteristic polynomial of $M$ for instance). Writing $p(z)=z^d-\sum_{i=1}^{d}a_{d-i} z^i$, from the expression $x_H(t)=P_H M^t x(0)$ valid in the mean-field model, it readily follows that
\begin{equation}
x_H(t)=\sum_{i=1}^{d} a_i x_H(t-i).
\end{equation}
In other words, the observables $x_H(t)$ follow an auto-regressive dynamics. Instead of estimating unknown coefficients of matrix $M$, one could therefore estimate directly the auto-regressive coefficients $a_i$. The latter approach is potentially simpler, and suffices if one is only interested in forecasting $x_H(t)$.  The coefficients $a_i$ however do not have a direct physical interpretation, in contrast to the coefficients of $M$. Estimation of $M$ is therefore better suited to assess the impact of specific measures.
\end{remark}

\paragraph*{Reaction to shocks}
If on given day $t_0$ (start of confinement say), the parameters $\alpha_i$ and $\alpha_a$ abruptly change, the dynamics from $t_0$ onwards would again be linear, with a modified matrix $M(\theta_1)$ with $\theta_1 = ( \alpha_{i,1},\alpha_{a,1})$ corresponding to the new contact rates. This suggests the following approach for inferring these parameters.

Estimate the state vector $x(t_0)$ as $C u$, for some constant $C = C(\theta_0)$ such that $C u_H=x_H(t_0)$, where $u_H$ is the projection of $u=u(\theta_0)$ on the first $h$ vectors of the canonical basis. 

Leverage then the transient behaviour of the post-shock dynamics as follows.  Making the dependency of the post-shock matrix $M(\theta_1)$ on the unknown $\theta_1$  explicit, estimate the unknown $(\theta_0, \theta_1)$ as the minimizers of the optimization problem:
\begin{equation}
\min_{\substack{\theta_0, \theta_1 \\ F(\theta_0)=\hat{\lambda}(t_0)}} \; \sum_{t> T_0} \bigg(\left[M(\theta_1)^{t-t_0} C(\theta_0)u(\theta_0)\right]_H -x_H(t) \bigg)^2.
\end{equation} 

\begin{remark}
 The above optimization criterion is essentially motivated by the assumption of constant contact parameters before and after the shock. This is a strong assumption, verified at best in an approximate sense. See \cite{Lavielle} which considers multiple phases in an SIR dynamics, to be fitted to observations. In Section \ref{section_numerical} we will consider two shocks.	 Time-varying contact rates will be discussed below in Section \ref{section_mobilite}.
\end{remark}

We just saw how to leverage Perron's theory  to perform inference in the mean-field version of the model. 
Let us now describe an inference method in the probabilistic model which -- as already explained -- is more relevant when dealing with small populations, e.g. when studying sub-critical evolutions in a decay phase, or spread relaunch phases.

\subsection{Kalman filtering for the branching process} 
In the following paragraph we describe an inference method in the probabilistic setting that enables to estimate the state $X(t)$ (including hidden trajectories) by updating the estimation at every new time step. The vector of average numbers per type at generation $t$, conditionally on the initial state to be $X(0) \in \mathbb{R}^J$, is $M^t X(0)$.

\paragraph*{Measurement errors} First, we describe a model for measures, that is the observed data $X^*_H$, as follows:
\begin{equation}
\label{eq:measures}
X^*_H(t) = P_H X(t) + v(t),
\end{equation} where $P_H$ is the $h \times J$ matrix of the projection on the $h$ last components of the canonical basis, and $v(t)$ is a centered noise, with covariance matrix $R(t)$, representing the confidence in measurements.

\paragraph*{Model errors} Now, we introduce the filtration 
$\cF_t := \sigma \left(X^*_H(0), \ldots, X^*_H(t) \right)$, that is the $\sigma-$algebra generated by all measurements up to time $t$.
We provide estimates of states $X(t)$, as well as the estimated errors. We will use the following notations:
\begin{itemize}
	\item $\hat{X}(t)$: an estimate of state $X(t)$ given $\cF_t$, and $P(t)$ its covariance matrix,
	\item $\hat{X}(t|t-1)$: an estimate of $X(t)$, given $\cF_{t-1}$, and $P(t|t-1)$ its covariance matrix.
\end{itemize}

At time $t$, we write $$X(t) = MX(t-1) + u(t),$$ with $u(t)=X(t) - MX(t-1)$ a centered vector. For $j \in [J]$, let $S_i$ denote the covariance matrix of vector of children from type $i$-parent. The covariance matrix $Q(t)$ of $u(t)$ can be computed as follows:
\begin{flalign*}
	Q(t)
	&:=\hbox{Var}\left( u(t) \right)\\
	&=\dE \left[\hbox{Var}\left(X(t) - MX(t-1) \big| X(t-1)\right)\right]+ \hbox{Var}\left( \dE \left[X(t) - MX(t-1) \big| X(t-1)\right]\right)\\
	&=\dE \left[\sum_{j \in [J]} X(t-1)_j S_j\right]+ 0\\
	&= \sum_{j\in[J]} (M^{t-1} X(0))_j S_j.
\end{flalign*} The fact that closed-form expressions for second moments are available is a consequence of the Markovian structure of our model. The same Markovian structure also implies that $u(t)$ is decorrelated from any $\cF_{t-1}-$measurable variable. 

\paragraph{Kalman filtering} The strategy is then to compute $\hat{X}(t|t-1)$ at each step, and to adjust the prediction with the measurement $X^*_H(t)$ to build $\hat{X}(t)$, assuming that  $\hat{X}(t)$ is of the form $$\hat{X}(t):= \hat{X}(t|t-1) + K(t) \left(X^*_H(t) - P_H\hat{X}(t|t-1)\right).$$  

Optimizing $K(t)$ in the quadratic error sense leads to \textit{Kalman filtering} (see e.g. \cite{grewal01}), which is the best linear recursive estimator in our setting, namely a hidden Markov model. With this method, at each time $t$, we perform the following predictions:
\begin{equation}
\begin{split}
\hat{X}(t|t-1) &= M \hat{X}(t-1),\\
P(t|t-1) &= M \cdot P(t-1) \cdot M^\top + Q(t),
\end{split}
\end{equation} then we adjust our estimates, updating with the new measure $X_H^*(t)$:
\begin{equation}
\begin{split}
K(t) &:= P(t|t-1) \cdot P_H^\top \cdot \left(P_H \cdot P(t|t-1) \cdot P_H^\top + R(t)\right)^{-1},\\
\hat{X}(t) &= \hat{X}(t|t-1) + K(t) \cdot (X_H^*(t) - P_H \hat{X}(t|t-1)),\\
P(t) &= \left(\mathrm{Id}_J - K(t) \cdot P_H \right) P(t|t-1) .
\end{split}
\end{equation}
The intermediate matrix $K(t)$ of size $J \times h$ is usually called the \textit{Kalman gain}. A possible initial condition consists  for instance in setting $X(0)=e_i$, that is one initial individual in a given state $i \in [J]$, with $P(0)=0_J$, the $J \times J$ null matrix.\\

The above discussion shows how to leverage the model structure to provide an alternative approach to maximum likelihood estimation approaches. 

Such techniques can be used in a single population scenario, as we illustrate in the next section. They could however become prohibitive for the multi-population scenarios we shall consider later, and for which the Kalman filtering approach may be an appealing alternative. 

\section{Numerical illustration on Paris hospitalization data}
\label{section_numerical}
In this section, we illustrate the probabilistic version of our model by fitting it to daily hospitalization data, and show how to use it to forecast future numbers of hospitalizations.

The population we will consider is that of department 75 (Paris, France). We still assume that the only observed trajectory is $X^{*}_H(t)$, the number of new hospitalizations, all other trajectories being hidden. A noisy version of this data, namely the number of new COVID-19 hospitalizations every day in Paris, can be found in public data provided by SurSaUD syndromic surveillance system. 

Though Kalman filtering approach -- as described in Section \ref{section_inference} -- would be natural here, we take a more direct, brute-force method, using Monte-Carlo simulation to infer the parameters.

\subsection{Simulation settings} For simulations of our probabilistic model, we take $p_i = 0.7$, $p_h = 0.05$ (see Transition probabilities in Section \ref{section_model}). We initialize the time range on February 8th, which is 18 days before the first local SARS-CoV2 death case in France. We focus on the pre-confinement and confinement phases, so the fitting is made up to May 5th, on a total of 72 days for the train set. The period from May 6th to May 12th is kept as the test set for prediction (see below).

Recall that for contact parameters, we assume $\alpha_i = \alpha_a = \alpha$. Furthermore we assume that it is piecewise constant on the time range, as follows: 
\begin{itemize}
	\item between day $0$ (February 8th) and day $t_2$, $\alpha=\alpha_1$.
	\item between day $t_2$ and day $t_3$, $\alpha=\alpha_2$.
	\item between day $t_3$ and day $T$ (May 5th), $\alpha=\alpha_3$.
\end{itemize}
The three phases can be identified to the \textit{pre-confinement phase}, the \textit{confinement phase I} and \textit{confinement phase II}, each phase being the result of several changes: mainly political decisions, but also work habits, observance of the rules, etc. In order to get a fitting as realistic as possible, the boundaries of these three phases are also inferred. 

\subparagraph{Inferred parameters, loss function} In total, six parameters $(X_E(0), \alpha_1, t_2, \alpha_2, t_3, \alpha_3)$ are tuned. $X_E(0)$ is the initial condition, that is the initial number of exposed individuals. This quantity is then divided and allocated to all variables $X_{E,d}(0)$ according to distribution $p_E$ described in Section \ref{section_model}.

Given $(X_E(0), \alpha_1, t_2, \alpha_2, t_3, \alpha_3)$, we sample several independent trajectories of $X(t)$. Now we consider two choices for the loss function.

We first define the $L^1$ loss function as follows:
\begin{equation}\label{eq_L}
\cL\left(X_E(0),\alpha_1,t_2, \alpha_2,t_3,\alpha_3\right) := \dE_{X_E(0)}\left[\frac{1}{T} \sum_{t=0}^{T} \left|X_H(t) - X^*_H(t) \right|\right],
\end{equation} where $X_H$ is the random trajectory under model $1$, and $X^*_H$ is the observed data. 

Second, since we expect exponential growth for the variables of interest (see e.g. \ref{subsection_branching_model} or \ref{subsection_perron}), we shall also consider the $L^1$-$\log$ loss function:
\begin{equation}\label{eq_L_tilde}
\widetilde{\cL}\left(X_E(0),\alpha_1,t_2, \alpha_2,t_3,\alpha_3\right) := \dE_{X_E(0)}\left[\frac{1}{T} \sum_{t=0}^{T} \left|\log X_H(t) - \log X^*_H(t) \right|\right].
\end{equation} 
For both choices, we aim to minimize the loss function on our train set. To do so, we perform a naive grid search in six dimensions, computing at each step an empirical expectation of the loss for the chosen parameters over sampled trajectories, in order to compute approximate solutions $(\hat{X}_E(0),\hat{\alpha}_1,\hat{t}_2,\hat{\alpha}_2,\hat{t}_3,\hat{\alpha}_3)$ of
\begin{equation}
\label{eq_minimisation}
\argmin_{\substack{X_E(0),\alpha_1,\alpha_2,\alpha_3 \in \mathbb{R}_+ \\ t_2, t_3 \in [T]}} \mathrm{Loss}\left(X_E(0),\alpha_1,t_2, \alpha_2,t_3,\alpha_3\right),
\end{equation} for $\mathrm{Loss} \in \left\lbrace \cL, \widetilde{\cL} \right\rbrace$.

\subparagraph{Prediction} After having estimated all the parameters, we use our model to predict the evolution of trajectories in a future time range. In our example, this time range runs from $T+1$ (May 6th) to $T_{\mathrm{pred}}$ (May 12th), that is a period of one week.

The performance of our predicition (over our test set) is evaluated in terms of the mean $L^1$ norm 
\begin{equation}\label{eq_L_pred}
{\cL_{\mathrm{pred}}} := \dE_{\mathrm{fit}}\left[\frac{1}{T_{\mathrm{pred}}-T} \sum_{t=T+1}^{T_{\mathrm{pred}}} \left| X_H(t) - X^*_H(t) \right|\right],
\end{equation} 
where the probability $\mathbb{P}_{\mathrm{fit}}$ is defined for previously inferred parameters $(\hat{X}_E(0),\hat{\alpha}_1,\hat{t}_2,\hat{\alpha}_2,\hat{t}_3,\hat{\alpha}_3)$ in equation \eqref{eq_minimisation}.  

We chose as our test criterion in \eqref{eq_L_pred} absolute deviation rather than deviation between logarithms because this corresponds to an error in the absolute numbers of hospital entrances, which has arguably a clearer operational meaning.

\subsection{Results} 
\begin{figure}
	\vspace{-1cm}
	\begin{subfigure}{\textwidth}
		\centering
		\includegraphics[width=\linewidth]{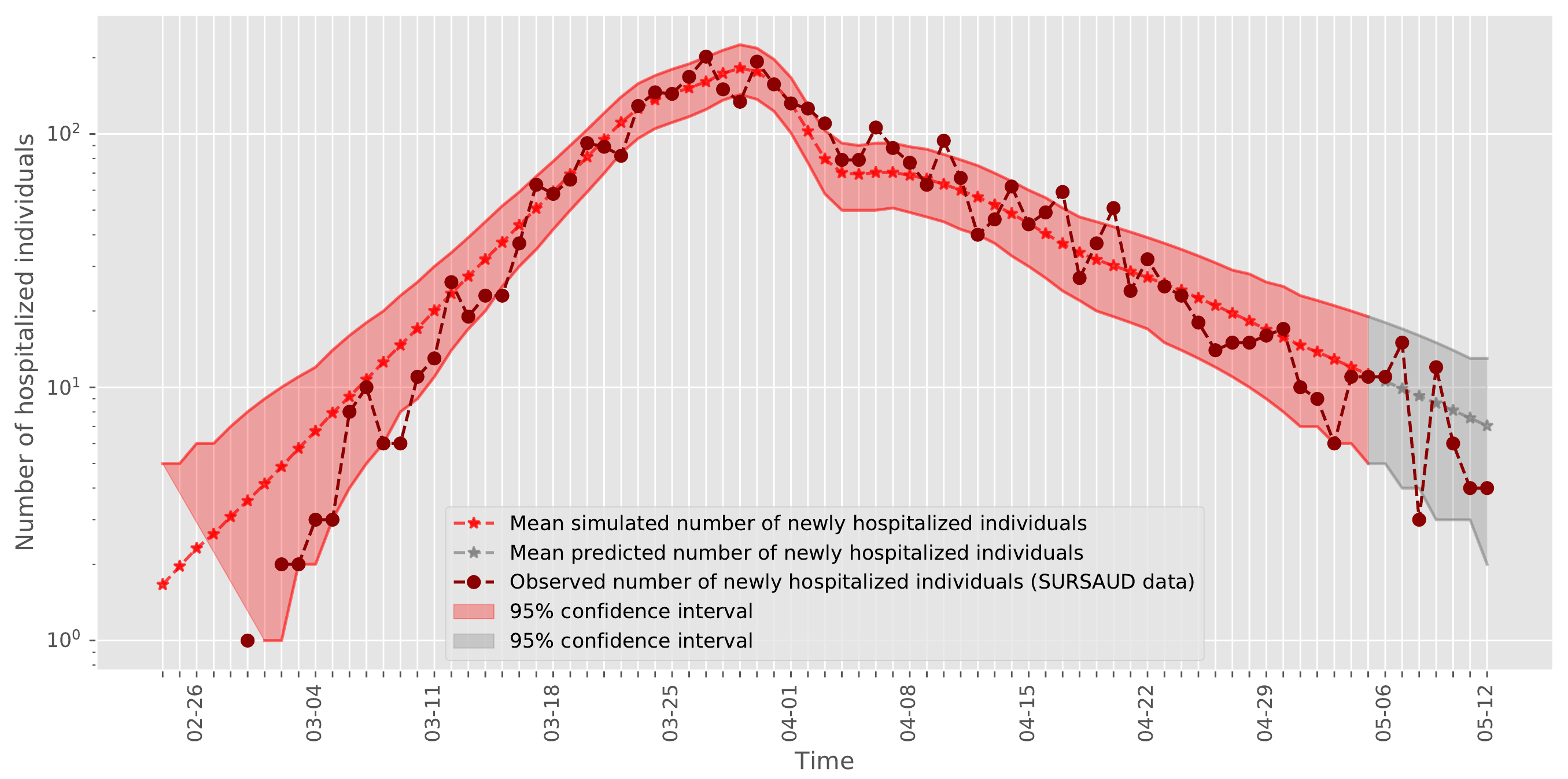}  
	\end{subfigure}
	\begin{subfigure}{\textwidth}
		\centering
		\includegraphics[width=\linewidth]{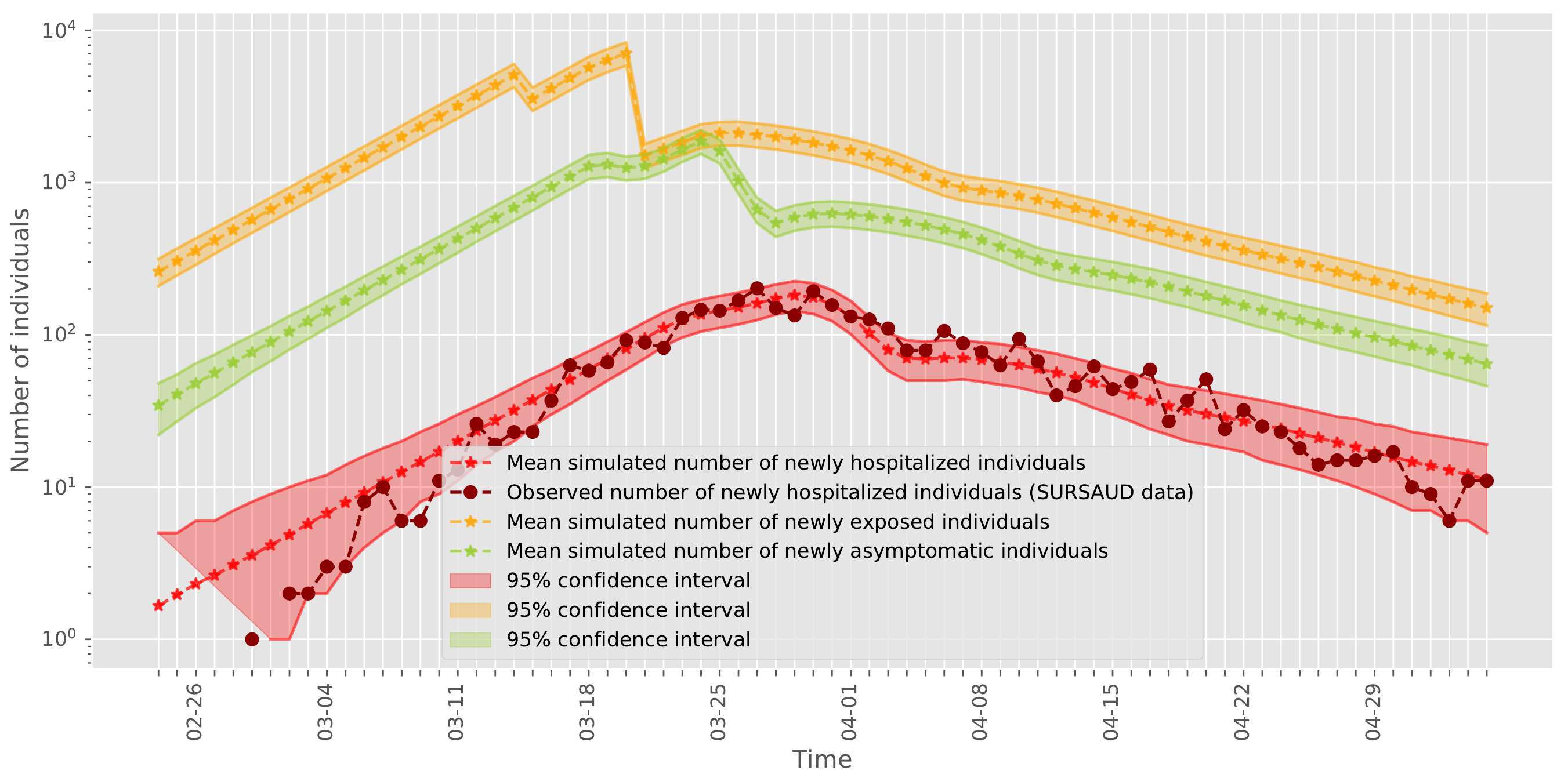} 
	\end{subfigure}
	\begin{subfigure}{\textwidth}
		\centering
		\includegraphics[width=\linewidth]{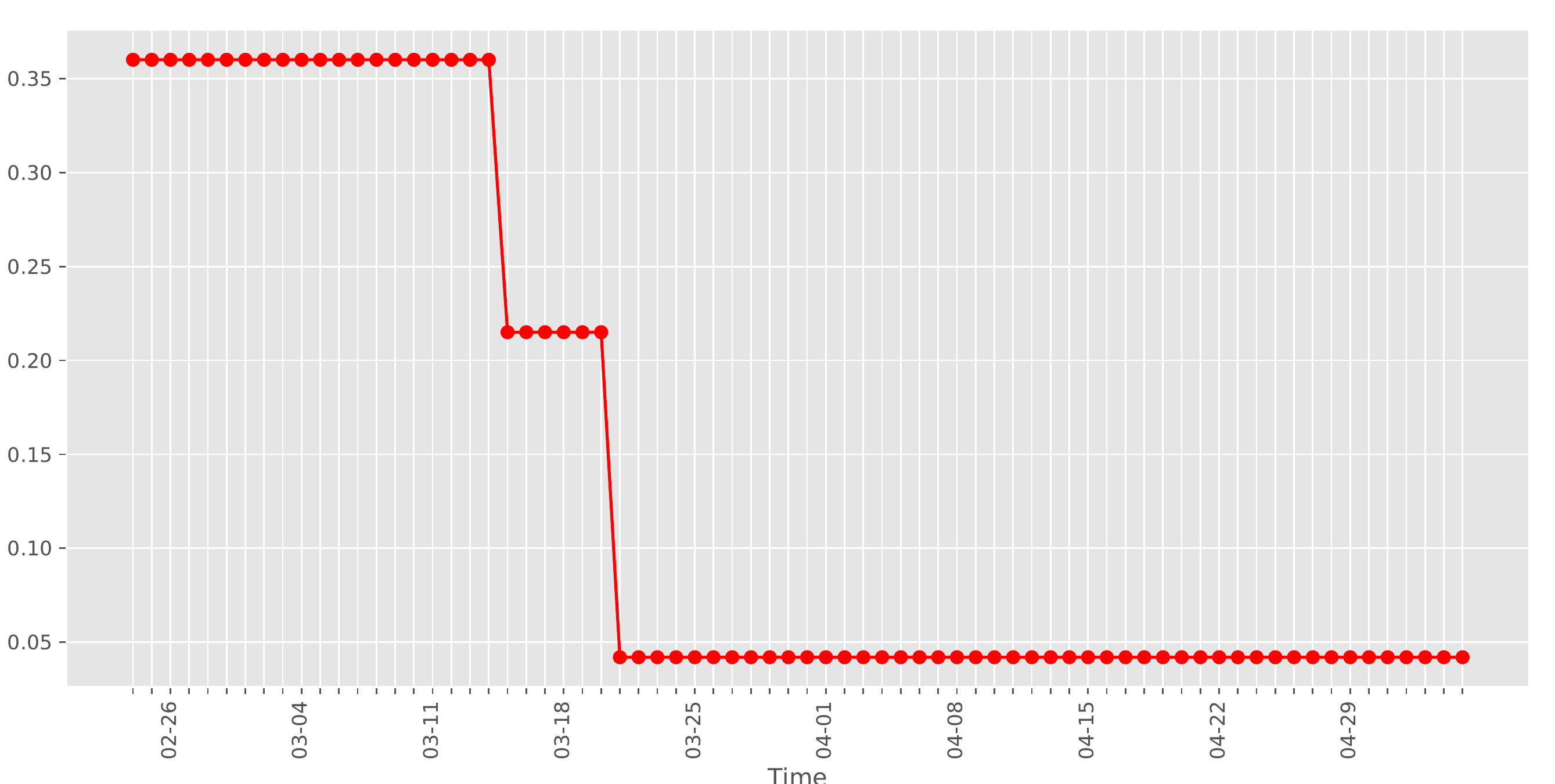}  
	\end{subfigure}
	\caption{Best fitting of our probabilistic model with SurSaUD data for Paris with $L^1$ cost function, with prediction, some hidden trajectories, and variation of contact rates $\alpha_a = \alpha_i$.}
	\label{fig_fits_sans_log}
\end{figure}

\begin{figure}
	\vspace{-1cm}
	\begin{subfigure}{\textwidth}
		\centering
		\includegraphics[width=\linewidth]{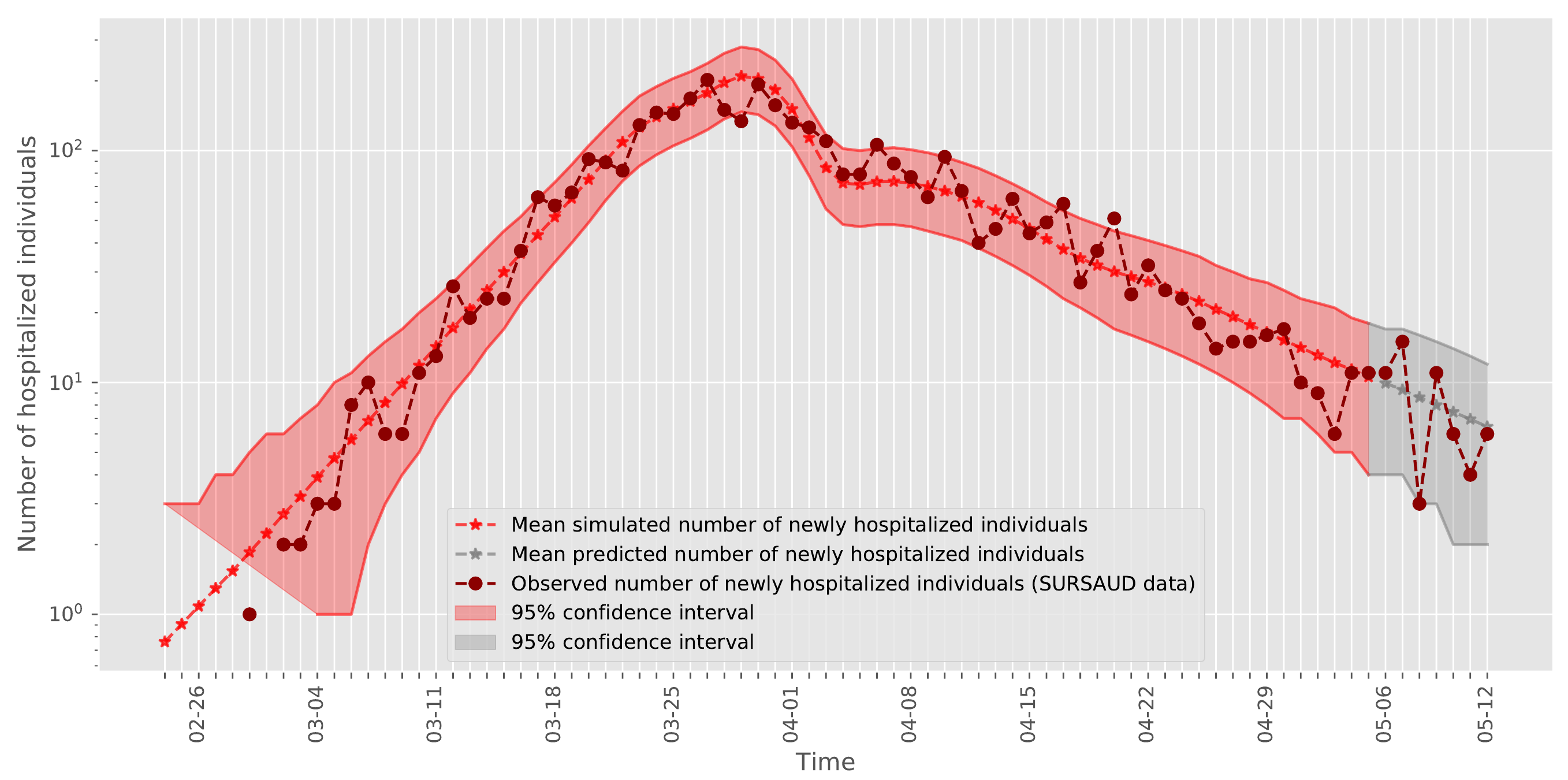}  
	\end{subfigure}
	\begin{subfigure}{\textwidth}
		\centering
		\includegraphics[width=\linewidth]{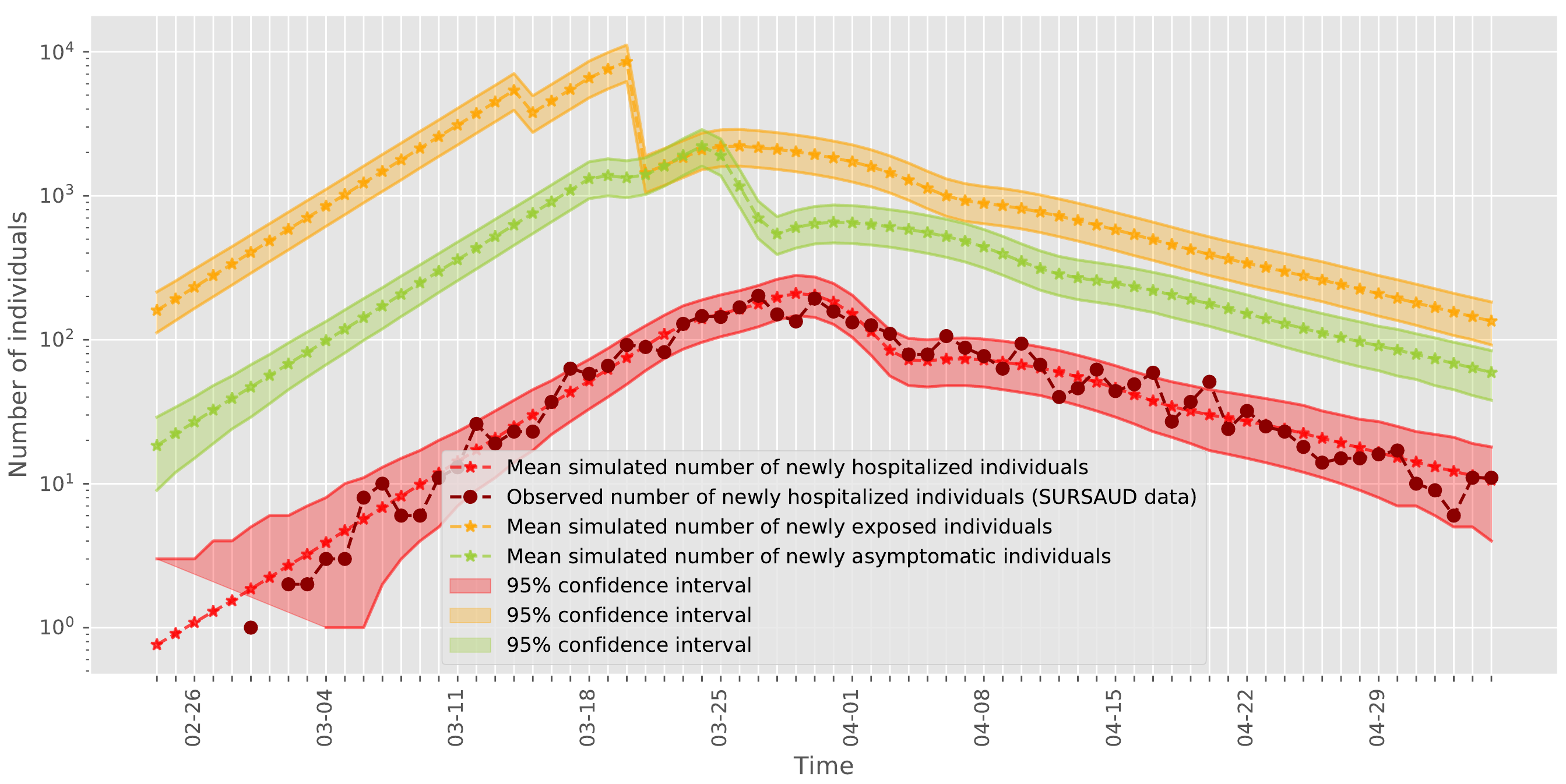} 
	\end{subfigure}
	\begin{subfigure}{\textwidth}
		\centering
		\includegraphics[width=\linewidth]{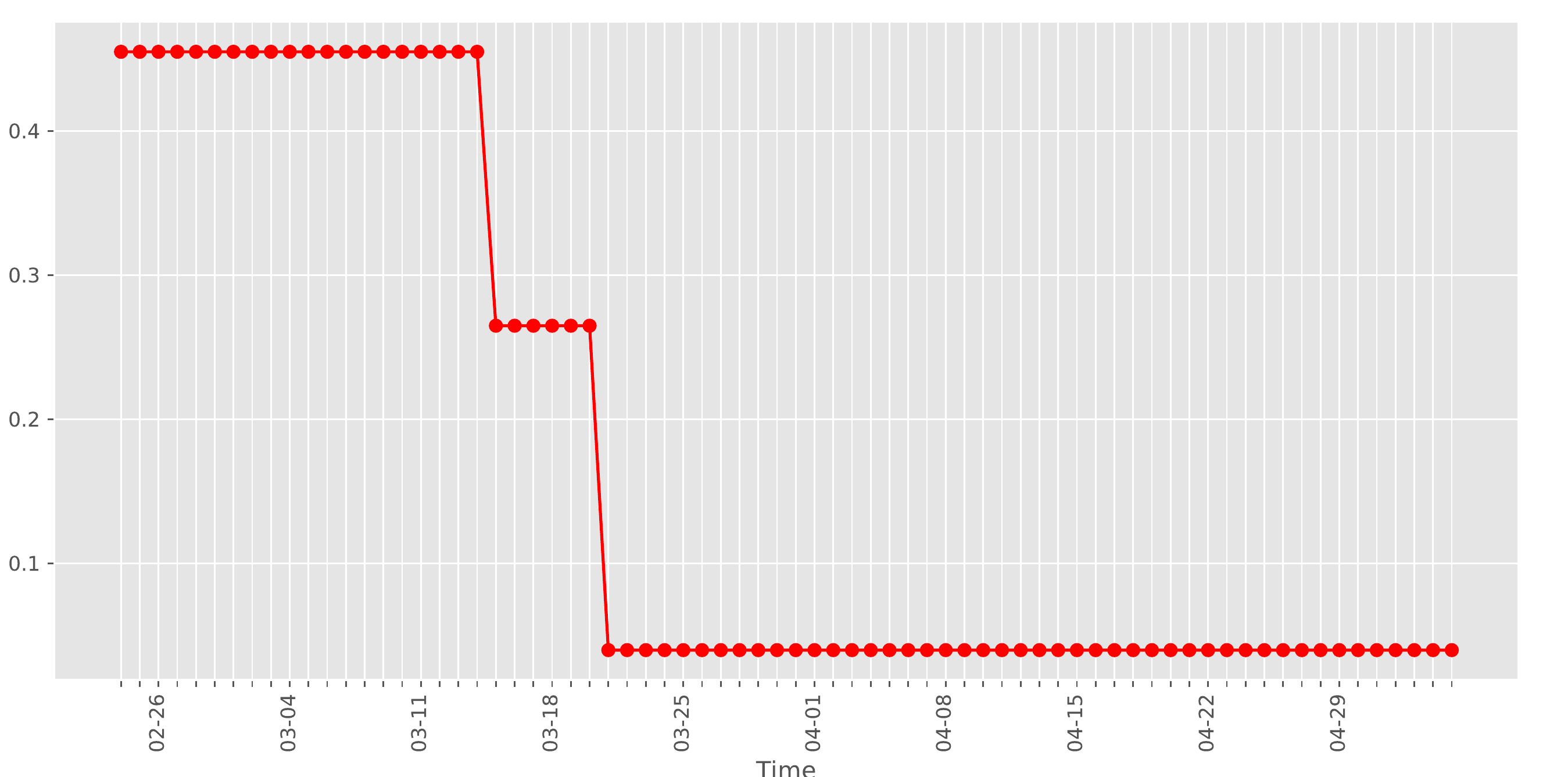}
	\end{subfigure}
	\caption{Best fitting of our probabilistic model with SurSaUD data for Paris with $\log$-$L^1$ cost function, with prediction, some hidden trajectories, and variation of contact rates $\alpha_a = \alpha_i$.}
	\label{fig_fits_avec_log}
\end{figure}

The inferred parameters with SurSaUD data, together with the training and test errors, are summed up in \Cref{table_simus}. Corresponding curves of inferred parameters, predicted hidden variables and hospital entrances are given in \Cref{fig_fits_sans_log} and \Cref{fig_fits_avec_log}.

\begin{table}[H]
	\begin{center}
		\leftskip -1.5cm
		\begin{tabular}{|c|c|c|c|c|c|c|c|c|}
			\hline
			\bf Loss function & $\bm{\hat{X}_E(0)}$ & $\bm{\hat{\alpha}_1}$ & $\bm{\hat{t}_2}$ & $\bm{\hat{\alpha}_2}$ & $\bm{\hat{t}_3}$ & $\bm{\hat{\alpha}_3}$ & \bf Fitting error & \bf Prediction error $\bm{\cL_{\mathrm{pred}}}$\\
			\hline
			$\bm{L^1 (\cL)} $ & $59$ & $0.360$ & 03-15 & $0.215$ & 03-20 & $0.042$ & $11.1195 \pm 0.0232$ & $ 4.0093 \pm 0.0185$  \\ 
			\hline
			$\bm{L^1}$-$\bm{\log (\widetilde{\cL})}$ & $18$ & $0.455$ & 03-15 & $0.265$ & 03-20 & $0.040$ & $0.29904 \pm 0.00062 $ & $3.7201 \pm 0.0125$   \\ 
			\hline
		\end{tabular}
	\end{center}
	\centering
	\caption{Approximate values of inferred parameters, for department 75, approximated with the probabilistic model, with SurSaUD data. All errors are given with their 95\% confidence interval obtained by Monte Carlo simulations.} 
	\label{table_simus} 
\end{table}

Several remarks can be made from these results:
\begin{itemize}
\item The $\cL_{\mathrm{pred}}$ prediction error is of order $4$, which is to be compared to the typical diameter of a $95\%$ confidence interval in the prediction time range in \Cref{fig_fits_sans_log} and \Cref{fig_fits_avec_log}, which is about $10$. We measure a prediction error which is less than half a typical uncertainty on a sampled stochastic trajectory $X_H(t)$ with fitted parameters. This supports the claim that a prediction error of order $4$ is the best possible with our model.
	
\item For both loss functions, the optimal transition points $t_1$ and $t_2$ are identified on March 15th and March 20th. Some other strategies could have been used to identify phases, e.g. by fitting a piecewise linear regression model on the observations $\log X_H^*(t)$, and reporting them earlier on the contact rates, according to the distributions of phase durations, as done in recent work by Gaubert et al. \cite{Gaubert20}. For the Paris population, the major political decisions and events occurred on March 12th (closure of all schools and universities), March 14th (closure of all non essential public places), March 15th (municipal elections) and March 17th (lockdown). These phases do not necessarily exactly coincide with the transition points, this can be caused e.g. by assumptions made for time parameters (see Section \ref{section_model}), and delays due to observance of the rules. In addition, we fixed three phases, but this number is arbitrary and could be discussed (see e.g. \cite{Lavielle}).

\item From the figures, the $L^1$-${\log}$ Loss function $\widetilde{\cL}$ seems to give better results for the fitting part (which is not surprising since we are in log scale), but this is also true for prediction, as presented in \Cref{table_simus}. This can justify the use of the $L^1$-${\log}$ Loss function $\widetilde{\cL}$ in future work.

\item These results, though based on noisy data and perfectible with other datasets or broader inference, validate our model: with few parameters, the evolution of epidemics can be described in a satisfactory way: with the $L^1$-${\log}$ Loss function for instance, the mean relative error is $\sim 12.1 \%$ (in logarithmic scale) for the trajectory of $X_H(t)$.
\end{itemize}

As shown in this section, the model of \Cref{fig_SEIAR_model_1} can be useful to describe the evolution of crucial variables such as the number of daily hospitalized individuals, but also show hidden trajectories, and predict hospital load on future days.\\ 

We considered  piecewise constant contact rates in Section \ref{section_inference} for inference purposes, and performed corresponding numerical illustration in the present Section. We shall now consider the case of contact rates that are no longer piecewise constant, being potentially dependent not just on ``shocks'' such as the start of confinement, but also on other covariates. For instance, regular day-to-day mobility or commuting flows of population could also impact contact rates, as we now investigate. 

\section{Variable contact rates}
\label{section_mobilite}
Contact rates are now assumed to depend on piece-wise constant rates depending on specific temporal phases, as described before, and daily variables that are responsible for additional variability. Among these daily variables, some are related to user mobility, and in particular on the daily outflow, that is the number of day-return trips outside the region of interest.   


The model we advocate is then exactly as in  Section \ref{section_model}, except that contact rates $\alpha_i$, $\alpha_a$ are now time-dependent.



\paragraph{Mobility-dependent contact rates}
Assuming that the daily outflows $f(t)$ (assumed to be normalized so as to have zero mean and standard deviation 1) are known, the contact rates $\alpha_a, \alpha_i$ are now mobility-dependent in the following way:
\begin{equation}
\label{eq_alpha_gen}
\begin{split}
\alpha_i(t) & = F_i \left( \phi(t), f(t)\right),\\
\alpha_a(t) & = F_a \left( \phi(t), f(t)\right).\\
\end{split}
\end{equation} In the above,  $\phi(t)$ is a discrete phase (e.g. $\phi(t) \in \left\lbrace 1,2,3\right\rbrace $ as in Section \ref{section_numerical}), and $F_i, F_a$ are fixed functions. 
These could be taken as logistic functions in the variable $f(t)$;
alternatively, one might take
\begin{equation}
\label{eq_alpha_pop}
\begin{split}
\alpha_i(t) & = \alpha_{i,\phi(t)} \left(1+ \gamma_i f(t) \right),\\
\alpha_a(t) & = \alpha_{a,\phi(t)} \left( 1+ \gamma_a f(t) \right).
\end{split}
\end{equation}
With this formulation, variables $\alpha_{i,\phi(t)}$  (resp. $\alpha_{a,\phi(t)}$) is the value of $\alpha_i$ (resp. $\alpha_a$) when the commuting flow is at its equilibrium, and  $\gamma_i, \gamma_a$ are non-negative parameters.

\Cref{fig_exemple_alpha_mob_75} below gives illustrates the evolution of the contact rates for Paris based on Equation \eqref{eq_alpha_pop}, where the observables $f(t)$ are obtained from SFR mobile operator data.
\begin{figure}[H]
	\centering
	\includegraphics[width=0.9\linewidth]{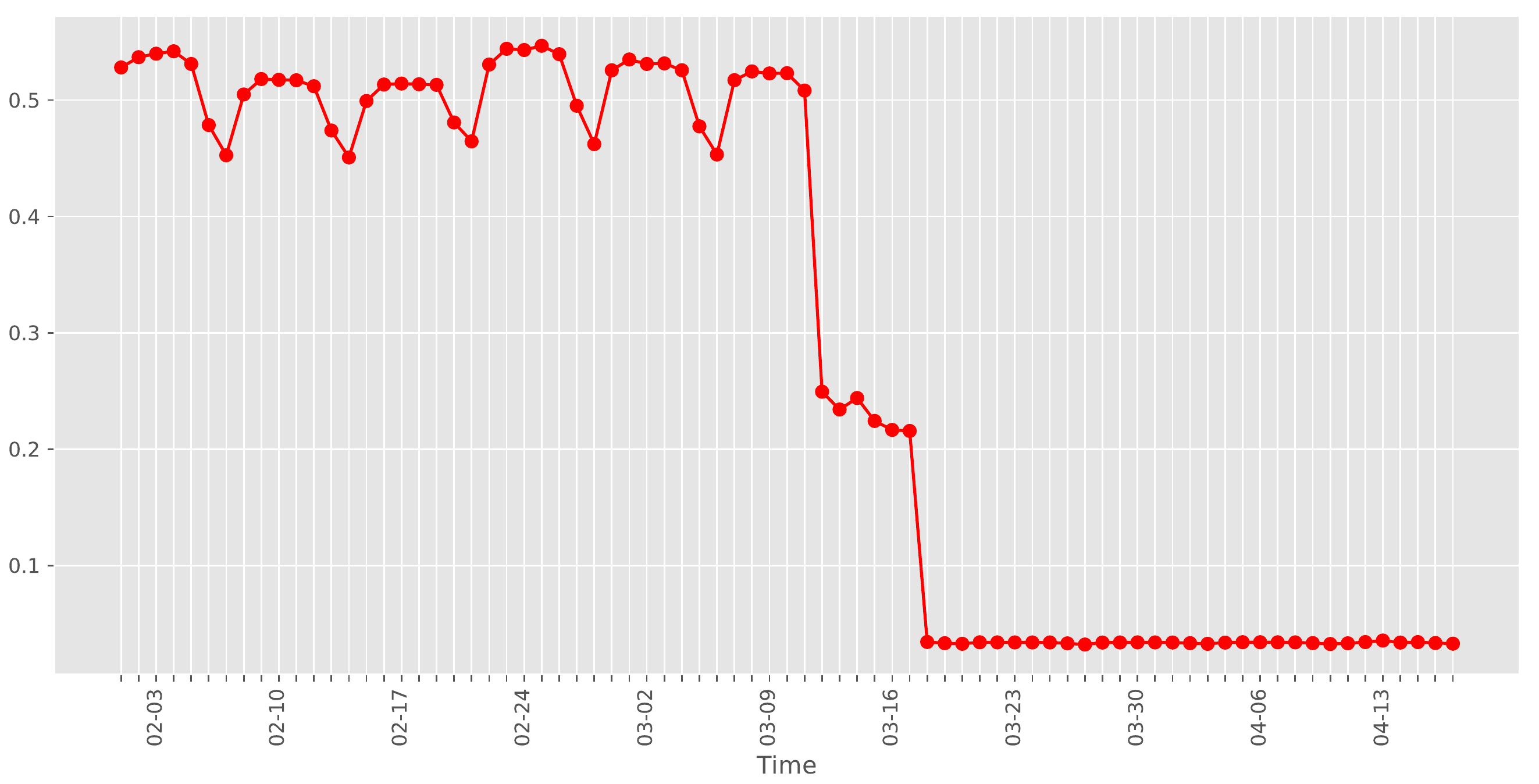}  
	\caption{Example of mobility-dependent contact rates with equation \eqref{eq_alpha_pop} for Paris, based on SFR data. Parameters: $\alpha_{i,1} =\alpha_{a,1} = 0.45$, $\alpha_{i,2} =\alpha_{a,2} = 0.3$, $\alpha_{i,3} =\alpha_{a,3} = 0.05$,  $\gamma_i = \gamma_a = 0.15$.}
	\label{fig_exemple_alpha_mob_75}
\end{figure}
The shape of the resulting contact rates reflects the discrete phases as before, together with  weekly periodicity patterns. Contact rates are lower on Saturdays and Sundays, when daily commuting trips are less numerous, as expected.\\

\begin{remark} In its mean-field version, the model with variable contact rates is 
$$
x(t+1)=\left[M_0+\sum_{k=1}^K \alpha_k(t) M_k\right]x(t),
$$ 
for fixed matrices $M_0,\ldots,M_K$ and scalars $\alpha_k(t)$ that depend on mobility covariates relative to time $t$. Assume for tractability that the covariates $\alpha_k(t)$ are small, so that heuristically we have the first order approximation, for $\tau\in[d]$, 
$$
x(t)\approx M_0^{\tau} x(t-\tau)+\sum_{i=1}^{\tau}\sum_{k=1}^K \alpha_k(t-i)M_0^{i-1}M_k  x(t-i).
$$
Following on Remark \ref{rem_ar}, let $p(z)=z^d-\sum_{i=1}^d a_i z^{d-1}$ be a degree-$d$ polynomial such that $p(M_0)=0$. We then have the first-order expansion in the $\alpha(t)$:

\begin{flalign*}
x(t) &\approx \sum_{i=1}^d a_i M_0^{d-i}x(t-d)+\sum_{i=1}^{d}\sum_{k=1}^K \alpha_k(t-i)M_0^{i-1}M_k  x(t-i)\\
&\approx \sum_{i=1}^d a_i\left[ x(t-i)-\sum_{j=1}^{d-i}\sum_{k=1}^K\alpha_k(t-i-j)M_0^{j-1}M_k x(t-i-j)\right] + \sum_{i=1}^{d}\sum_{k=1}^K \alpha_k(t-i)M_0^{i-1}M_k  x(t-i).
\end{flalign*}

Provided that for all $i=0,\ldots,d-1$ and all $k\in [K]$, there exist matrices $N_{k,i}$ and polynomials $p_{k,i}(z)=\sum_{r=0}^{d(k,i)}b_r(k,i)z^r$ of degree $d(k,i)\le i$ such that $P_H M_0^iM_k=N_{k,i}P_H p_{k,i}(M_0)$, then up to terms of first order one has
$$
P_H M_0^iM_kx(s)=N_{k,i}\sum_{r=0}^{d(k,i)}b_r(k,i) P_H M_0^r x(s)\approx N_{k,i}\sum_{r=0}^{d(k,i)}b_r(k,i) x_H(s+r).
$$
Plugged into the previous display, this entails that the observations $x_H(t)$ satisfy an auto-regressive relation with coefficients involving the covariates $\alpha$, of general form:
\begin{equation}
x_H(t)=\sum_{i,j=1}^{d}\alpha(t-i)P_{i,j} x_H(t-j),
\end{equation}
where $P_{i,j}$ are constant matrices.
A possible approach could then consist, for prediction purposes, in fitting such an auto-regressive relation while learning suitable functions $\alpha(t)$ of available covariates. 
\end{remark}

Besides commuting outflows mentioned above, another aspect of mobility  that plays an important role is that of {\em routing}, that is change of residence location: this happens during holiday periods, but has also happened at the start of lockdown, where a significant exodus from dense urban areas towards countryside has been observed. 

We now describe an extension of our model that captures such routing.

\section{Model with Routing Mobility}
\label{section_routing} 
To introduce routing, we must first consider an extension of our basic model to cover multiple cohorts, or sub-populations.  For the sake of readability we only describe the generalized model in its mean-field version.  However a probabilistic version could easily be described along the same lines as in the single population case. 

We thus introduce a subdivision of the global population into cohorts $c \in \cC$. These could for instance be given by a couple $(r,a)$ characterizing geographical regions $r$ and age ranges $a$. 

A potential alternative to the above definition of cohort $c$ is to define sub-population according to a given age range $a$, the region $r$ in which the corresponding persons usually live, and the region $r'$ in which they slept the previous night. In that case, we have $c=(r,r',a)$. This notion of cohort is particularly important to describe our routing model, and account for viral dissemination through regions.

We denote by $\cN_c(t)$ the size of cohort $c$ at time $t$. 


\subsection{Model parameters for subpopulations}
The model is as follows. Each cohort $c \in \cC$ has its own biological parameters $\theta_c$ (encapsulating e.g. contact rates, distributions of phase durations, etc.). A natural assumption is to take these to depend only on age range. Per age parameters could be obtained from hospital-collected statistics.
\paragraph{Routing dynamics of $\cN_c(t)$}
For a given cohort $c \in \cC$, the time evolution of $\cN_c(t)$ is the result of population flows. We represent such flows in the following manner: we have
\begin{equation}
\cN_c(t+1)=\sum_{c' \in \cC}R_{c',c}(t) \cN_{c'}(t) + \cE_c(t+1),
\end{equation}where $R_{c',c}(t)$ is the fraction of individuals counted in $\cN_{c'}(t)$ on day $t$, that have migrated to be counted in $\cN_{c}(t+1)$ on day $t+1$. The case $c'=c$ of sedentary individuals (no migration) is also taken into account. New external arrivals in $c$ (say, from a foreign country) are accounted for by $\cE_c(t+1)$. We also assume that for all cohort $c'$,
\begin{equation}
\sum_{c \in \cC} R_{c',c}(t)\le 1,
\end{equation} allowing for departures outside the considered population (say, abroad) when the sum is strictly less than $1$.

\paragraph{Daily activity, contact between sub-populations} 
On  day $t$, a contact intensity factor $n_{c,c'}(t)$ gives the average number of individuals from cohort $c'$ that a typical individual from cohort $c$ will encounter. It is natural to assume that 
\begin{equation}\label{eq_infection_rate_c0}
\cN_{c}(t)n_{c,c'}(t)=\cN_{c'}(t)n_{c',c}(t)
\end{equation}
Parameter $q_{c,c'}$ represents the probability that upon a contact between individuals from population $c$ and population $c'$, infection gets propagated from that in $c$ to that in $c'$. It is natural to assume that $q_{c,c'}$ only depends on the age ranges of $c$ and $c'$. We then let 
\begin{equation}
\label{eq_infection_rate_c}
\alpha_{c,c'}(t):=q_{c,c'} n_{c,c'}(t), 
\end{equation} the infection rate from $c$ to $c'$ on day $t$. For simplicity we do not distinguish between asymptomatic/prodromic (previously $\alpha_a$) and symptomatic individuals (previously $\alpha_i$) here. 

Note that the above dynamics, and in particular Equations \eqref{eq_infection_rate_c0}, \eqref{eq_infection_rate_c} generalize the mobility-dependent contact rates of Section 4 to the multi-subpopulation case.

\subsection{Routing step in the mean field version}
Each sub-population $c$ on day $t$ is characterized by state vector $x^c(t)$, given by concatenation of vectors $x^c_E$, $x^c_P$, $x^c_{I_1}$, $x^c_A$, $x^c_{I_2}$, $x^c_H$. As before this is a vector of size $5h +1$, $h$ being the maximal number of days each phase lasts.

To specify the transition between day $t$ and day $t+1$ we consider the following two-stage mechanism: a first stage consists in an operation similar to the previously considered dynamics, extended to account for multiple cohorts $c$, giving vector $y(t)$ as a linear transform of vector $x(t)$. A second stage representing the routing of individuals between cohorts gives $x(t+1)$ as a linear transform of $y(t)$. Specifically we have 
\begin{equation}
y^c(t)=N_{c,c}(t) x^c(t)+\sum_{c'} \alpha_{c',c}(t) M_{c,c'}x^{c'}(t),
\end{equation} where matrices $N=\left(N_{c,c}\right)_{c,c' \in \cC}$ and $M=\left(M_{c,c}\right)_{c,c' \in \cC}$ are characterized by the equations
\begin{flalign*}
\forall \tau \in \left\lbrace E,P,I_1,I_2,A \right\rbrace, \;  y^{c}_{\tau,d+1}(t)&= x^{c}_{\tau,d}(t)(1-r_{\tau}(d)),\\
y^{c}_{E,1}(t) &=\sum_{c' \in \cC} \alpha_{c',c} (t) \sum_{\delta>0}\left[(x^{c'}_{I_1,\delta}+x^{c'}_{I_2,\delta})(t)+ (x_{A,\delta}^{c'}+ x_{P,\delta}^{c'})(t) \right] ,\\
y^{c}_{P,1}(t)&=\sum_{\delta>0} x^{c}_{E,\delta}(t) r_E(\delta),\\
y^{c}_{I_1,1}(t)&=p_{i}\sum_{\delta>0}x^{c}_{P,\delta}(t)r_P(\delta),\\
y^{c}_{A,1}(t)&=(1-p_{i})\sum_{\delta>0}x^{c}_{P,\delta}(t)r_P(\delta),\\
y^{c}_{I_2,1}(t)&=(1-p_{h})\sum_{\delta>0}x^{c}_{I_1,\delta}(t) r_{I_1}(\delta),\\
y^{c}_{H}(t)&=p_{h}\sum_{\delta>0}x^{c}_{I_1,\delta}(t)r_{I_1}(\delta).
\end{flalign*} 
The routing stage is then provided by:
\begin{equation*}
\begin{pmatrix}
x^c_E\\
x^c_P\\
x^c_{I_1}\\
x^c_A\\
x^c_{I_2}
\end{pmatrix} (t+1) = 
\sum_{c'}R_{c',c}(t)
\begin{pmatrix}
y^c_E\\
y^c_P\\
y^c_{I_1}\\
y^c_A\\
y^c_{I_2}
\end{pmatrix}(t).
\end{equation*}

\subsection{Estimation of routing fractions and numbers of contacts}
\paragraph{Observation/estimation of routing fractions} An ideal scenario is when routing fractions $R_{c,c'}(t)$ are directly available. However, this is typically not  the case. We therefore describe a more plausible situation, for which we also propose a potential approach for estimating these fractions.


Assume that  $c=(r,r',a)$, $r$ representing the usual address of individuals and $r'$ the place where they spent the previous night. Assume that we observe the quantities $\cN_c(t)$, as well as, for all pair of regions $(r_1,r_2)$,  the quantities:
\begin{equation}
\Delta_{r_1,r_2,a}(t):=\sum_{r,r'} \cN_{(r,r_1,a)}(t) R_{(r,r_1,a),(r',r_2,a)}(t),
\end{equation} that is the flow of people of age range $a$ who slept on night before $t$ in $r_1$ and slept on the following night in $r_2$. The usual address of these individuals is however assumed unknown.

In such a situation, we propose the following approach, popular in the literature on traffic matrices for management of communication networks. Estimate the unobserved quantities $R_{c,c'}(t)$  as the solution $\widehat{R}_{c,c'}(t)$ of 
\begin{subequations}
	\begin{alignat}{2}
	&\!\max_{R_{c,c'} \geq 0}        &\qquad& \sum_{c,c'} R_{c,c'}\ln\left(\frac{1}{R_{c,c'}}\right)\\
	&\text{subject to} &      & \forall c', \; \sum_{c} R_{c',c}\le 1\\
	&                  &      & \forall (r_1, r_2), \; \sum_{r,r'} \cN_{r,r_1,a}(t) R_{(r,r_1,a),(r',r_2,a)}=\Delta_{r_1,r_2,a}(t).
	\end{alignat}
\end{subequations}This is a maximum entropy criterion. As a concave maximization program, it can be solved with efficient numerical methods.

\paragraph{Estimation of mean numbers of contacts $n_{c,c'}(t)$}
Contacts counted in the contact intensity factor $n_{c,c'}(t)$ may occur in many circumstances. Among these we distinguish contacts at home, at work, at school, in transit (public transportation) as parameters on which preventive measures can be brought to bear. We may split such contacts accordingly, writing
\begin{equation}
n_{c,c'}(t)=n^H_{c,c'}(t)+n^W_{c,c'}(t)+n^S_{c,c'}(t)+n^T_{c,c'}(t).
\end{equation}
Let us consider estimation of $n^T_{c,c'}(t)$. Assume we have access to additional observations about the mobility of individuals of all sub-populations. For instance, we may have statistics showing that on day $t$, approximately $n_{c,z}(t)$ individuals of population $c$ have visited a location $z$. We may represent by $z$ a time-space location, e.g. being in site $s$ during a given hour $h$ of the day.
\begin{remark}
	The variables $n_{c,z}(t)$ could be unprocessed counts observed on day $t$. It may however be more appropriate to apply preliminary filtering on such unprocessed counts (e.g. applying a weighted averaging of raw counts over a past time window) to improve estimation quality of contact rates.
\end{remark}
We may then let
\begin{equation}
\cN_c(t) n^T_{c,c'}(t)=\sum_z \beta_z n_{c,z}(t) n_{c',z}(t),
\end{equation}
where the parameters $\beta_z$ are to be inferred, and capture the density of contacts to be expected in (time-space) location $z$. In this formulation, the parameters to be inferred are: parameters $\beta_z$, and infection probabilities $q_{c,c'}$. This model is over-parameterized, since multiplying the $\beta_z$ by $\eta$ and dividing the $q_{c,c'}$ by $\eta$ leaves the model unchanged for all $\eta \ne 0$. This can be easily solved, forcing for instance the parameters $\beta_z$ to have mean $1$. While this over-parameterization is easily circumvented, it remains a challenge to effectively fit all the free parameters of the above model, as well as to define the proper spatio-temporal granularity associated with the time-space locations $z$. 

\section*{Conclusion}
In this study we laid the foundations of a general model of COVID-19 epidemics which captures user mobility and contact tracing. We described it as a multi-type branching process, proposed inference methods to perform short-term prediction, that we illustrated on the Paris hospitalization data. We also introduced extensions to capture contract tracing and case isolation, and finally proposed extensions to capture user mobility, distinguishing between commuting and routing mobility.

Future work will further exploit this model together with hospital incidence and mobility data to forecast epidemic progress and assess impact of mobility on infectious contacts.

\paragraph{Acknowledgements:} The authors are grateful to SFR for making available to them time series of human mobility across departments in France. This data, partially illustrated by Figure \ref{fig_exemple_alpha_mob_75}, has inspired the design of the models described in the last two sections. 
%
%
%
\newpage
\bibliographystyle{plain}
\bibliography{covid,graphical_combined,bib_2,BibCommunityDetection}

\end{document}